\documentclass[11pt]{article}

\usepackage[all=normal,paragraphs,wordspacing,tracking]{savetrees}
\usepackage{times}
\usepackage{setspace}
\usepackage{amsthm}
\usepackage{amsmath}
\usepackage{amsfonts}
\usepackage{amssymb}
\usepackage{graphicx}
\usepackage{pdfpages}
\usepackage{stmaryrd}
\usepackage{enumitem}
\usepackage{nicefrac}
\usepackage{verbatim}
\usepackage[normalem]{ulem}
\usepackage{verbatimbox} 
\usepackage{bm}
\usepackage{xfrac}

\usepackage[utf8]{inputenc}
\usepackage{amssymb}
\usepackage[colorinlistoftodos]{todonotes}
\usepackage{algorithm}
\usepackage[noend]{algpseudocode}
\usepackage{enumerate}
\usepackage{xcolor}
\usepackage{framed}
\usepackage{pifont}
\usepackage{multicol}
\usepackage{lipsum}
\usepackage{tikz}
\usetikzlibrary{calc}
\usepackage{wrapfig}
\usepackage{array}
\usepackage{relsize}
\usepackage{comment}
\usepackage[title]{appendix}
\usepackage{caption}
\usepackage{subcaption}
\usepackage[margin=1in]{geometry}
\usepackage{hyperref}

\newtheorem*{notation*}{Notation}

\newcommand{\FC}{{\sc{FC}}} 
\newcommand{\VC}{{\sc{VC}}} 
\newcommand{\RpH}{{\sc{RpH}}} 

\newcommand{\dmi}{\Delta m_i} 
\newcommand{\smart}{smart mining}
\newcommand{\smarter}{smarter mining}

\newcounter{linecounter} 
\newcommand{\linenumbering}{\ifthenelse{\value{linecounter}<10} 
	{(0\arabic{linecounter})}{(\arabic{linecounter})}}

\renewcommand{\thelinecounter}{\ifnum \value{linecounter} >  
	9\else 0\fi \arabic{linecounter}} 

\newlength {\squarewidth}

\title{Mind the Mining}
\author{Guy Goren\\
	Technion IIT\\
	\texttt{sgoren@campus.technion.ac.il}
	\and
	Alexander Spiegelman\\
	VMware Research\\
	\texttt{spiegelmans@vmware.com}
}

\begin{document}

\date{}

\maketitle

\begin{abstract}

In this paper we revisit the mining strategies in proof of work
based cryptocurrencies and propose two strategies, we call
\emph{smart} and \emph{smarter mining}, that in many cases strictly
dominate honest mining.
In contrast to other known attacks, like selfish mining, which induce
zero-sum games among the miners, the strategies proposed in this
paper increase miners' profit by reducing their variable costs (i.e.,
electricity).
Moreover, the proposed strategies are viable for much smaller miners
than previously known attacks, and surprisingly, an attack performed by
one miner is profitable for all other miners as well.

While saving electricity power is very encouraging for
the environment, it is less so for the coin's security.
The smart/\smarter\ strategies expose the coin to under 50\%
attacks, and this vulnerability might only grow when new miners join
the coin as a response to the increase in profit margins induced by these
strategies.

%
%

\end{abstract}





\thispagestyle{empty}

\newpage

\setcounter{page}{1}

\section{Introduction}
\label{sec:Intro}

As of the end of 2018, the total cryptocurrency market cap is above
100 Billion dollars. Dozens of new coins emerge every month and the
industry of digital mining is blooming.
According to~\cite{MarketState}, the vast majority
of the coins are based on the PoW technology~\cite{back2002hashcash},
which received a lot of attention with the introduction of
Bitcoin~\cite{Bitcoin}.
The main idea is that a lot of power has to be wasted in order to
change the coin state, what makes Sybil attacks impossible, and thus makes consensus possible in an anonymous open networks.
The drawback of this technology is the huge amount of electricity
it consumes. 
As of 2018, the Bitcoin alone consumes more electricity
than 159 countries including Nigeria and
Morocco~\cite{Powercompare}.
A part from not being environmental friendly, the huge waste of
power induces very high costs on coins maintenance.

The main entities in a cryptocurrency system are the miners. 
They maintain the state and preserve the security by doing work that
requires a lot of power, and in return they get to mint new
coins.
For economical and security reasons, cryptocurrencies try to
enforce a fixed rate of new minted coins\footnote{Presently, for example, the
Bitcoin systems~\cite{Bitcoin} tries to enforce miners to collectively mint
75 new coins every hour.}.
This is done by determining how much power has to be invested by a
miner in order to mint one coin, which is usually called the
\emph{difficulty}.
Ideally, if the total mining power (by all miners) invested in a coin
would have been known at any given time, then the difficulty would
have been accurately calculated and the fix minting rate could be
enforced.
However, this is not the case. 
Miners can freely join and leave the system at any time, and free to stop mining if it is not profitable for them.

\begin{figure}[th]
    \centering
     \begin{subfigure}[t]{0.49\textwidth}
        \centering
        \includegraphics[width=3.2in]{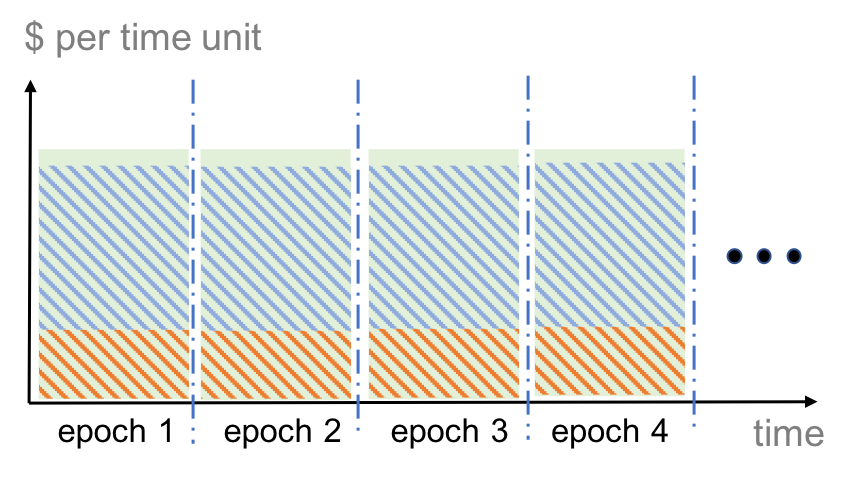}
        \caption{Desired equilibrium: Never idle.}
         \label{}
     \end{subfigure}%
    ~
     \begin{subfigure}[t]{0.49\textwidth}
        \centering
        \includegraphics[width=3.2in]{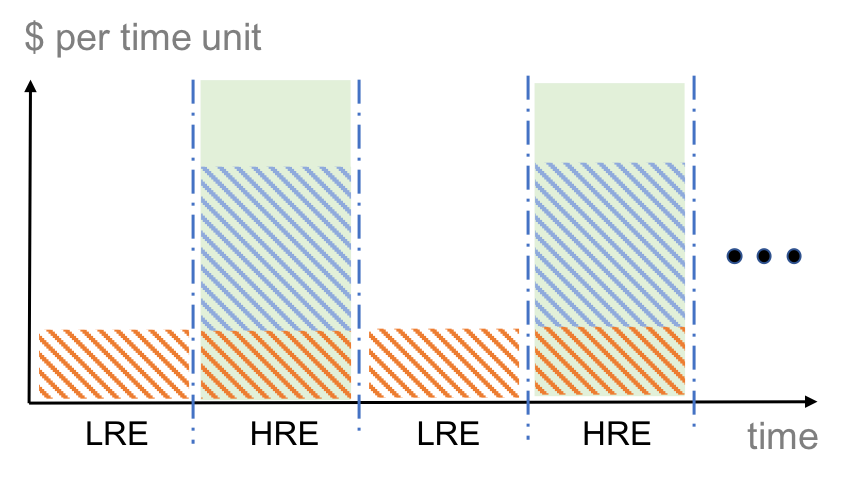}
         \caption{Smart mining: Idle every other epoch.}
    \end{subfigure}
    \vskip\baselineskip
    \centering
    \begin{subfigure}{1\textwidth}
        \centering
        \includegraphics[width=4in]{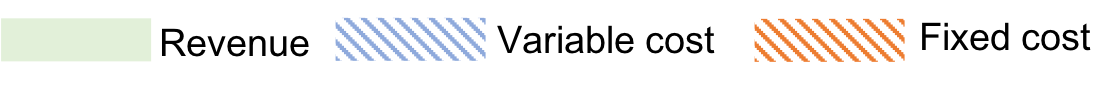}
        \label{}
    \end{subfigure}

    \caption{A miner's profit in an epoch is the revenue minus the
    costs, which is the uncovered green area.
    In the desired equilibrium, the profit is the same in
    every epoch.
    In smart mining, the miner pays the fixed cost and gains no profit
    in LRE epochs, but gain large profit in HRE epochs. 
    If the profit bonus in HRE epochs is bigger than the fixed cost
    plus the epoch profit in the desired equilibrium strategy, then
    the smart mining strategy is more profitable for the miner.}
    \label{fig:concept}
\end{figure}

Therefore, the best cryptocurrency systems can do is to predict the
future based on the estimation of the total mining power from the
past.
This is done by dividing executions into epochs, where each epoch
consists of a fix number of minted coins.
(For example, in Bitcoin~\cite{Bitcoin}, each epoch consists of
2016 block, each of which mints 12.5 coins - as of 2018.) 
The difficulty for
each epoch is calculated based on the estimation of the total mining power in the previous epoch.
This means that coins can adjust to changes in the total mining power
only during epoch changes, which makes them vulnerable to sudden
changes in total mining power.
This vulnerability was already noticed before as one that can lead to
a problem called ``blockchain death spiral''~\cite{spiral}, in which
miners suddenly leave a coin (possibly because its value suddenly
dropped), leaving it with high difficulty and forcing a long epoch.
This can lead to serious throughput decrease in strong coins,
and to a total death of small ones.
In this paper we further explore this vulnerability and show how
miners con exploit it for their benefit.
In particular, we show that the ``desired equilibrium'' in which 
miners always mine with their total power is not an equilibrium.
surprisingly, we show that in many cases stop mining and being idle
is a strictly better strategy.
The basic concept is based on the fact that the total mining power in
each epoch determines the difficulty, and thus also the revenue of
miners, in the next epoch.
Therefore, if a miner does not mine during an epoch, it loses the
revenue of this epoch, but it saves the cost of the power in
this epoch and gain more revenue in the next epoch due to the
difficulty adjustment.%
\footnote{In practise there is a maximum factor
by which the difficulty can change between two consecutive epochs.
However, it does not invalidate our attack, it only adds another
parameter for our analysis, which for simplicity and readability we
choose to omit in this paper.}
We call the strategy in which a miner alternately mine in an epoch
and then idle in the following epoch a \emph{smart mining} strategy,
where epochs in which the miner mines and epochs in which it does not
mine are called \emph{high revenue epochs (HRE)} and \emph{low revenue
epochs (LRE)}, respectively.
See Figure~\ref{fig:concept} for illustration.
Interestingly, the benefit of the smart miner strategy does not come
on account of other miners. 
On the contrary, other miners befit from it even more since they lose
nothing in low revenue epochs and gain in high revenue
epochs.
Having this in mind, we show that in some cases miners
benefit the most from mining with only part of their mining power.
We call this strategy \emph{smarter mining}.
The smarter mining strategy can be seen as an optimization of smart
mining. 

Note that while smart and smarter mining is a win-win for all miners
and the environment, the coin security is compromised during low
revenue epochs.
Not only that a smart miner does not mine in low revenue epochs, it
might be the case that other miners gain from joining him and thus
leaving the coin exposed to attacks (during low revenue epochs) by a
bad miner that controls less than 51\% of the total mining power.
The smart and smarter mining strategies increase the profit margins in
high revenue epochs, which might bring new miners to the coin
and potentially correct the coin security.
However, surprisingly, we get exactly the opposite effect. 
When new miners join the high revenue epochs, the difficulty in the
low revenue epoch goes higher, and thus the revenue per time unit in the low revenue
epochs decreases, which might force miners to leave these epochs and
expose the coin to even more attacks.

Analysing the strategic behavior of miners in cryptocurrency systems
has become a subject to a large study in the last few
years~\cite{liao2017incentivizing, SelfishMining,
OptimalSelfishMining, StubbornMining, carlsten2016instability,
MiningGames, Johnson14, MinersDilemma, Schrijvers16,
hellemans2019mining, spiegelman2018game, tsabary2018gap,
bonneau2016buy}.
The pioneering ``Selfish-Mining" attack strategy demonstrated
that deviating from the mining protocol can be beneficial even
without a majority of the mining power~\cite{SelfishMining}.
Their strategy, however, requires 25\% or more of the mining power,
which is relatively high.
Smart mining, on the other hand, is relevant for smaller miners as
well. For example, if the fixed costs represent 10\% of the miner's
total costs, having just 12\% of the mining power suffices.
Moreover, the long term effect of allowing new players to join the
system exposes significant differences.
While in selfish mining the joining of new players restores the
system's security, in smart mining the opposite happens.
Players that join the coin for economical reasons, unintentionally
further damage the coin security and increase its vulnerability to
under 51\% attacks.

In~\cite{carlsten2016instability}, Carlsten et al. showed that selfish
mining can be made profitable for a miner with a low hash power share
in a model in which miners are getting paid by transactions fees
rather than by minted coins.
In the same model, Tsabary and Eyal~\cite{tsabary2018gap}, showed
that miners can increase their profit by not mine (being idle), and
thus reduce their electricity costs, when the total fees amount of
available transactions is low.
In this paper we consider the more standard model, that is currently
used in practice, in which miners are paid in new minted coins, and
to the best of our knowledge, we are the first to propose a
dominating strategy that is beneficial to all miners (i.e., the
attacker and honest miners), but decreases the coin security.

The rest of the paper is organized as follows:
In section~\ref{sec:PoW} we give an overview on PoW-based
cryptocurrencies and in section~\ref{sec:Model} we define our model.
In Section~\ref{sec:RM} we introduce and analyse the smart and
smarter mining strategies, and in Section~\ref{sec:BestResp} we
analyse the other miners best response and discuss the implication on the coin
security.
In Section~\ref{sec:Disc} we conclude the paper.


\section{Proof of work overview}
\label{sec:PoW}

In the next section we define a model that aims to capture the core
of the proof of work (PoW) based cryptocurrencies (e.g.,
Bitcoin~\cite{Bitcoin} and Ethereum~\cite{ethereum}) mechanism that
we need to demonstrate our attack.
In this section we give a short and simplified description of how it
works.
PoW, which was first introduced in Bitcoin~\cite{Bitcoin}, is a novel
approach to solve randomized synchronous anonymous byzantine leader
election.
The idea is that there is a known to all puzzle which all parties
try to solve by performing hash operations, where the difficulty of
the puzzle determines the probability for a single hash operation to
solve the puzzle.
The first party that maneges to solve the puzzle broadcasts the
solution, and as a result it is elected as the leader.
Parties are usually called miners, the number of hash functions that
a miner can perform in a time unit is called mining power.
An important property of PoW, which make it useful for
cryptocurrencies, is the fact that the probability of a miner to be
the first to solve the puzzle and become the leader is equal to the
ratio of the miner's mining power out of the total mining power of
all miners.

Most of the PoW-based cryptocurrency systems use PoW in the following
(simplified) way. 
They start from a known-to-all genesis block that determine the first
puzzle.
Once a solution to the $k^{th}$ puzzle
is found, the
chosen leader (the party that found the solution) broadcasts it to
all parties.
The solution forms a block that is added to the block-chain, which in
turn determines the next puzzle.
However, this basic idea has several challenges that needed to be
addressed:
\begin{itemize}
  
  \item First, the system must give incentives for
  the miners in order to encourage them to participate in the protocol (e.g., solve
  the puzzles).
  To this end, miners get paid, by new coins they mint, when they
  find solutions.
  
  \item Second, a crucial requirement from a PoW-based cryptocurrency
  is that solutions are broadcast faster than they are
  found~\cite{decker2013information}.
  This is essential in order to reduce the possibility of disagreement
  on the leader (i.e., forks) -  
  if a miner finds a solution first, but another miner find a
  different solution before the first solution was broadcasted, then we
  cannot know who is the true leader.
  To overcome this problem (by reducing the probability of such
  event), cryptocurrencies try to control the expected rate in which
  solutions are found.
  Recall that the difficulty determines the probability of a single
  hash to solve the puzzle, and since the probability of every hash is
  independent form the other hashes, if the total mining power is
  known, then difficulty can be set to determine the expected rate of
  solutions.
  
  \item The third challenge is how to estimate the total mining power.
  Recall that in public cryptocurrencies, miners can leave and join
  the system whenever they want, so there must be a dynamic mechanism to track these changes and adjust the difficulty accordingly.
  In most of the PoW-based cryptocurrencies it is done in the following way.
  The execution is divided into epochs, where each epoch consists of  
  a fixed number $B$ of blocks (puzzle solutions), e.g., $B=2016$ in
  Bitcoin. When epoch $ep_i$ is over, the system uses the real time it took for
  epoch $ep_i$ to complete in order to estimate the total mining power
  used during this epoch.
  Then, this estimation is used to calculate the new difficulty for
  the next epoch.
  
\end{itemize}
\noindent
Note that since the difficulty and the reward (number of new minted
coins) for finding a solution is always known, a miner can calculate
the expected revenue it gets for every hash it performs, and by taking into account its costs, the miner can
estimate the expected profit per hash.
The desired equilibrium in PoW-based cryptocurrencies, which is
important in order to reason about security, is that all miners
always (during all epochs) mine with their total mining power.
In this paper we consider the standard demand and supply economic
assumption in the desired equilibrium, by which the profit of the
miners are negligible, and show that miners have strictly better
strategies that compromise the coin security.
It is important to note that the better strategies we demonstrate in
this paper exists also for an arbitrary $\epsilon$ profit, with only
minor changes in the numerical results.

\section{Model and Definitions}
\label{sec:Model}
For simplicity of analysis, we define a deterministic model that
captures the core of PoW-based cryptocurrencies.
Our model does not use puzzles and a difficulty to determine
the probability of a single hash to solve the puzzle. Instead, we
deterministically define the revenue each miner gets from a single hash
operation and how many hashes needed in total to complete an epoch.
Note that by defining a deterministic model we give the system more
control and thus our results apply for the real probabilistic case as
well.
Our model consists of a single coin $C$ and a set of miners $\Pi =\{
p_1,\ldots,p_n\}$, which mine for $C$ by performing hash operations.
Each miner~$p_i$ possesses a hashing power that enables it to
perform $m_i$ hashes per time unit, and we allow miners to choose
when to hash.
We denote by $M \triangleq \sum_1^n m_i$ the total hash power all
miners collectively posses.  
We assume that miners have fixed and variable costs, that is, a miner
$p_i$ pays a fixed price $FC_i$ every time unit regardless of how
many hashes it performs, and an additional variable price $VC_i$ per
every hash.

Recall that a PoW-based cryptocurrency progress in epochs, where each
epoch consists of a fix number~$B$ of blocks, and the system sets
the difficulty in order to control the expected time $T$ of the
epoch.
This is done by choosing the difficulty of the puzzles in a way that
requires $T\cdot M$ total hash operations in \emph{expectation} in order to
solve $B$ puzzles.
In our model we straighten the coin by allowing to define the required
number of hash operations in every epoch deterministically.
An execution in our model progresses in sequential epochs
$ep_1,ep_2,\ldots$, where each epoch $ep_k$ consists of $H_k$ hashes
performed by all miners.
That is, epoch $ep_k$, $k>1$, starts immediately after $H_{k-1}$
hashes were collectively performed during epoch $ep_{k-1}$.
We denote by $t_k$ the number of time units it took for epoch $ep_k$
to complete.
Initially, $H_1 = M \tau$, where $\tau$ is a system parameter.
Intuitively, $\tau$ is the desired duration of time the coin wants
every epoch to be.
Note that if all miners mine during the first epoch, then the epoch
duration is exactly $\tau$ time units, i.e., $t_1= \tau$.
As for the next epochs, for every $k>1$, $H_k =
\frac{H_{k-1}}{t_{k-1}}\tau$.
Similar to a real system, $\frac{H_{k-1}}{t_{k-1}}$ estimates the
total mining power used during epoch $k-1$, and $H_k$ is calculated so
that if the mining power stays the same during epoch $k$, its duration
will be the desired $\tau$.

In a real system, a miner can estimate, by the difficulty, its
expected revenue and profit from every hash it performs. 
Here we define it deterministically.
Recall that in a real system a miner gets to mint a fixed number of
coins for every solution, and thus the system ``pays'' a fixed total
of rewards in every epoch.
Here, since we deterministically define the number of hash operations
in every epoch, we can deterministically define the revenue a miner
gets for every performed hash.
Let $w$ be the total reward the coin $C$ divide during an
epoch.
(Again, for example, Bitcoin~\cite{Bitcoin} pays 12.5 Bitcoins
per solution, so for an epoch of 2016 blocks, Bitcoin~\cite{Bitcoin}
pays 25200 Bitcoins.)
For every $k \geq 1$, the revenue per hash in epoch $ep_k$ is $RpH_k
\triangleq \frac{w}{H_k}$ for every miner.

The miners in our model are rational in a way that they try to
maximize their profit over time.
Therefore, they may choose not to utilize there full mining
capabilities at all times.
However, for simplicity, we assume that miners do not change their
power during an epoch. 
We denote by $\hat{m_i}[k] \leq m_i$ the number of hash operations
per time unit miner $p_i$ performs during epoch~$ep_k$, and the cost
per-time-unit of miner $p_i$ at epoch~$ep_k$ by $C_i[k] \triangleq FC_i
+ VC_i\hat{m_i}[k]$.
The revenue per time unit of miner $p_i$ during epoch $ep_k$ is
denoted by $R_i[k]\triangleq\frac{w}{H_i}\hat{m_i}[k]$, and
the profit per-time-unit by $P_i[k]\triangleq R_i[k] -
C_i[k]$.
The utility function of a miner $p_i$ is defined as the
average profit per unit of time over an unbounded execution:
\[
u_i \triangleq \lim_{K \to \infty} \frac{\Sigma_{k=1}^K P_i[k]\cdot
t_k}{\Sigma_{k=1}^K t_k}
\]

\begin{figure}[th]
    \centering
        \includegraphics[width=4in]{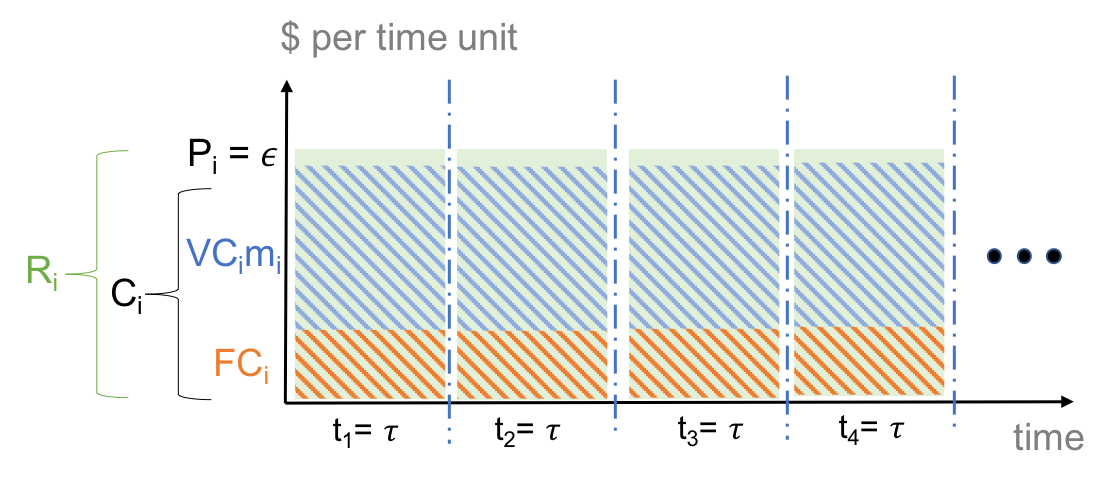}
        \caption{Demonstration of the definitions over the ``desired
        equilibrium'' strategy.
        The epoch parameter $k$ is omitted.}
    \label{fig:model}
\end{figure}

We assume that miners are economical beings that would not mine for a
loss, but would join mining if it is profitable.
Thus, by the classical model of supply and demand \cite{Wealth1776, ricardo1817}, we assume that the profit of miners in the desired
equilibrium, in which miners mine with there full capacity, is some $\epsilon\ge0$.
Figure~\ref{fig:model} uses the ``desired equilibrium'' to demonstrate
our definitions.
To capture this in our model we set $w = \tau \Sigma_{i=1}^n (FC_i + VC_im_i + \epsilon)$.
For simplicity and readability we set~$\epsilon=0$\footnote{In classical economic theories of free markets \cite{Wealth1776, ricardo1817}, the revenue and costs strive for equality, resulting in a negligible~$\epsilon$. This is not fundamental for the smart mining strategy, and only slightly alters the numerical results.}.
Intuitively, $w$, the total reward the coin divides during an epoch, is set to be equal to the total cost the miners pay if they mine in full capacity during the epoch.


\section{Rational Mining}
\label{sec:RM}
Recall that the ``desired equilibrium" in a PoW-based cryptocurrency
mining system, is such that all miners mine with full capacity in all
epochs.
However, due to the difficulty update mechanism this ``desired
equilibrium'' is actually not an equilibrium in many cases, and a
miner~$p_i$ has a strongly dominant strategy diverting from the
protocol.

In principle,~$p_i$ can increase the revenues in the next epoch by mining less in the current epoch.
This results in~$p_i$ losing revenues in the current epoch but also in~$p_i$ reducing its costs.
Therefore, opening a possibility for increased profits.
We call the consequent mining strategy \smart.

\subsection{Smart Mining}
\label{sec:smart}
Assume that all miners are mining and the system has achieved its ``desired equilibrium'' and remains stable during all epochs~$k`<k$.
Hence, according to the protocol design (and intention), for every
$k'<k$ the values of $t_{k`}, H_{k'}$ and \RpH$_{k`}$ are fixed to
$\tau, M\tau$ and $\frac{w}{M\tau}$ respectively.
We now show a strategy by which a miner~$p_i$ can benefit by
diverting from the protocol.
The strategy is for~$p_i$ to stay idle (not mine) during
epochs~$\{ep_k, ep_{k+2}, ep_{k+4},...\}$ and to mine with full power
during epochs~$\{ ep_{k+1}, ep_{k+3}, ep_{k+5},...\}$.
Lets analyze~$p_i$'s profits.

\paragraph{Epoch~$k$.} Since $t_{k-1}=\tau$, we have in $ep_k$ that
$H_k= M\tau$ as well. If~$p_i$ remains idle during $ep_k$, then
\[t_k=\frac{H_k}{M-m_i}=\frac{\tau}{1-\frac{m_i}{M}}.\]
Consequently, $p_i$'s profit for~$ep_k$ is $(-$\FC$_i\cdot t_k)$, which is negative if there are any fixed costs.

\paragraph{Epoch~$k+1$.} Since $t_k=\frac{\tau}{1-\frac{m_i}{M}} >
\tau$, the difficulty adjustment mechanism reduces the difficulty
which in turn increases the reward per hash and correspondingly also the profit per time unit. The resulting values are:
\begin{align*}
	H_{k+1}&= \frac{\tau}{t_k}\cdot H_{k}
	= \frac{M-m_i}{M} \cdot M\tau \\
	\mbox{\RpH}_{k+1} &= \frac{w}{H_{k+1}}
	= \frac{w}{(M-m_i)\tau}\\
	t_{k+1}&= \frac{H_{k+1}}{M}
	= \frac{M-m_i}{M} \cdot \tau \\
	P_i[k+1]&= m_i\mbox{\RpH}_{k+1} - \left(\mbox{\VC}_i  m_i+\mbox{\FC}_i\right)
	= \frac{m_i w}{(M-m_i)\tau} - \left(\mbox{\VC}_i  m_i+\mbox{\FC}_i\right)
\end{align*}

\begin{figure}[th]
    \centering
        \includegraphics[width=6.3in]{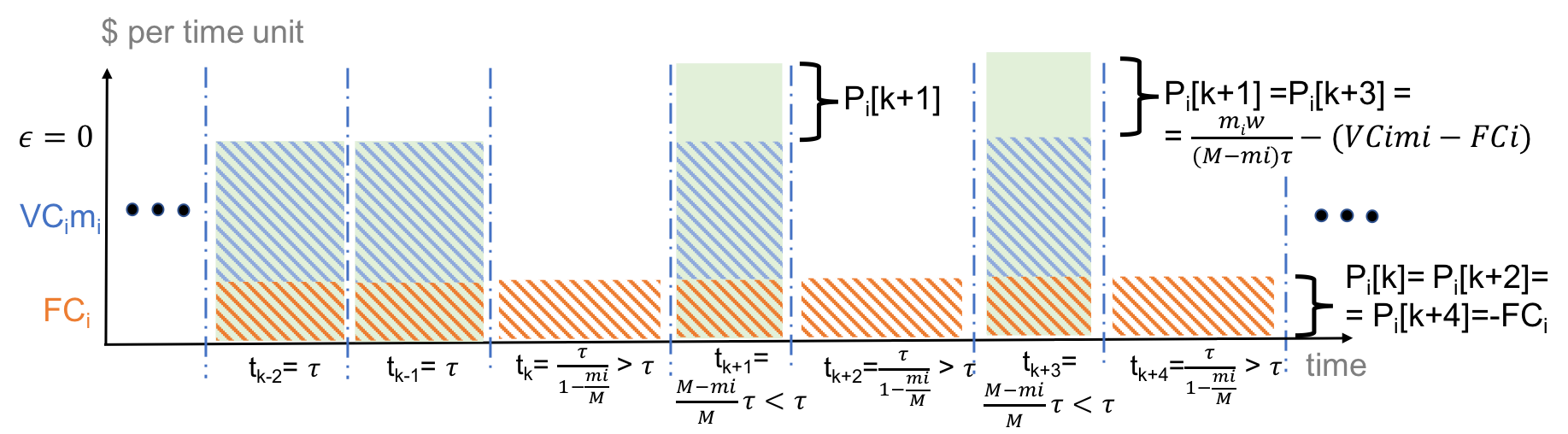}
        \caption{Smart mining strategy for a party $p_i$. Profitable
        when $P_i[k+1]\cdot t_{k+1} > P_i[k]\cdot t_k$.}
    \label{fig:smart}
\end{figure}

\paragraph{Epochs~$\ge k+2$.} Since $t_{k+1}=\frac{M-m_i}{M} \cdot \tau$ and $H_{k+1}= \frac{M-m_i}{M} \cdot M\tau$, we have in $ep_{k+2}$ that $H_{k+2} = M\tau$ as in $ep_k$. The rest is identical to $ep_k$.
Inductively, from here on epochs $\{ep_{k+2}, ep_{k+4},...\}$ result in the same values as in $ep_k$, and epochs $\{ep_{k+3}, ep_{k+5},...\}$ result in the same values as in $ep_{k+1}$.
Figure~\ref{fig:smart} illustrates the smart mining strategy.

\begin{figure}[h]
\centering
	\includegraphics[width=0.66\linewidth]{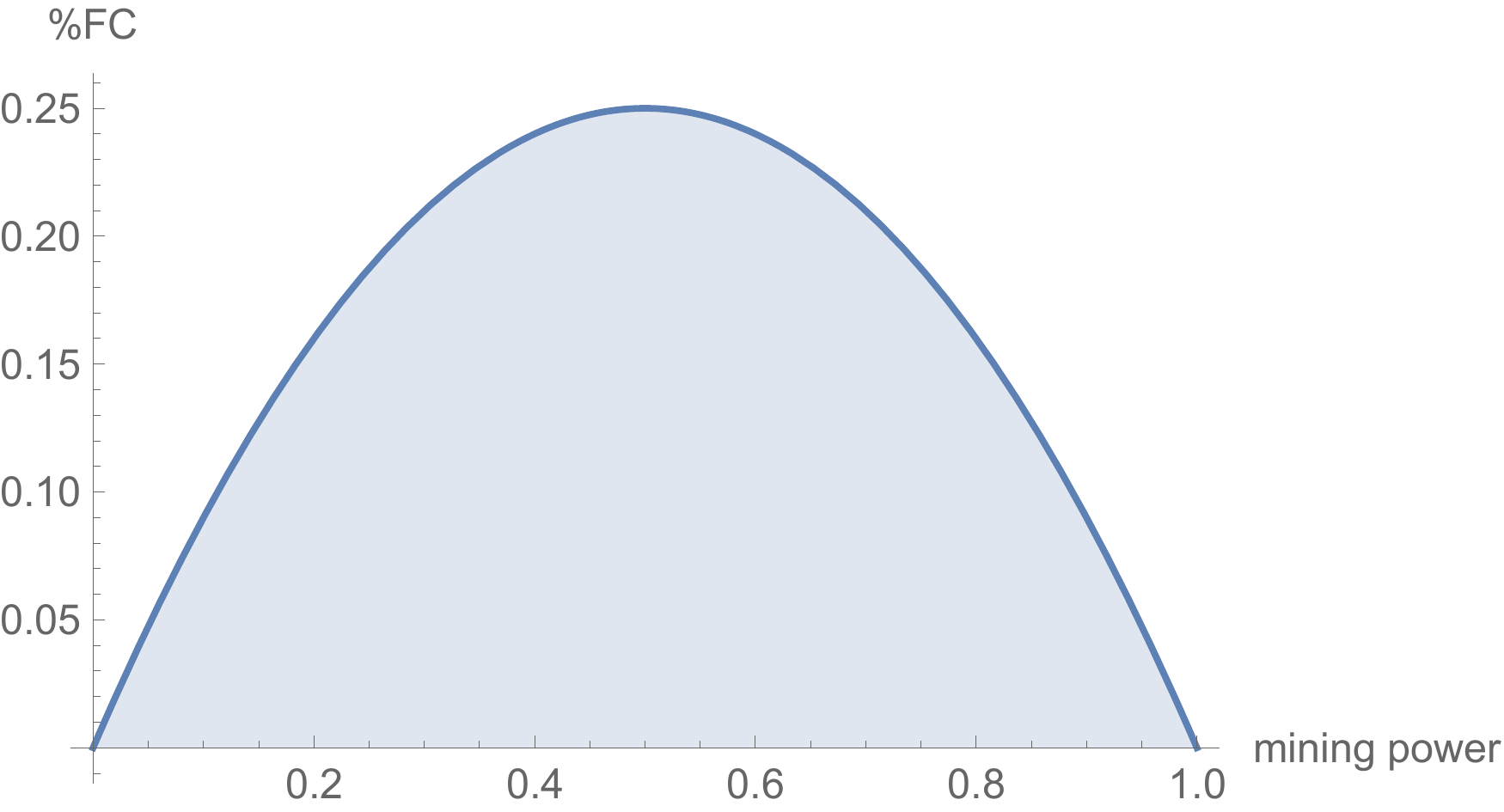}
	\caption{\smart\ strictly dominate honest mining in the area under the curve.
		$x$-axis:~$\frac{m_i}{M}$, $y$-axis:~$\frac{\mbox{FC}_i}{\mbox{\VC}_i \cdot m_i+\mbox{\FC}_i}$.}
	\label{fig:(1-x)x}
\end{figure}

\paragraph{}We are now ready to calculate $i$'s profit averaged over time as defined by the utility function.
\begin{align*}
	u_i &\triangleq \lim_{K \to \infty} \frac{\Sigma_{k'=1}^K
	P_i[k']\cdot t_{k'}}{\Sigma_{k'=1}^K t_{k'}}\\
	&= \lim_{K \to \infty} \left[ \frac{\Sigma_{k'=1}^K (P_i[2k'-1]\cdot
	t_{2k'-1}) + \Sigma_{k'=1}^K (P_i[2k']\cdot t_{2k'})}{\Sigma_{k'=1}^K t_{k'}} \right] \\[3ex]
	&= \frac{P_i[k+1]\cdot t_{k+1} + P_i[k]\cdot t_{k}} {t_{k+1} + t_k}\\
	&= \frac{ \left( \frac{m_i w}{(M-m_i)\tau} - \left(\mbox{\VC}_i m_i+\mbox{\FC}_i\right) \right) t_{k+1} -\mbox{FC}_i t_{k}} {t_{k+1} + t_k}\\
	&= \frac{ \left( \frac{m_i w}{(M-m_i)\tau} - \left(\mbox{\VC}_i m_i+\mbox{\FC}_i\right) \right) \frac{M-m_i}{M} -\frac{M}{M-m_i}\mbox{FC}_i } {\frac{M-m_i}{M} + \frac{M}{M-m_i}}\\
	&= \frac{  \frac{m_i w}{M\tau} - \left(\mbox{\VC}_i m_i+\mbox{\FC}_i\right) \frac{M-m_i}{M} -\frac{M}{M-m_i}\mbox{FC}_i} {\frac{M-m_i}{M} + \frac{M}{M-m_i}}
\end{align*}
In the ``desired equilibrium'' of the protocol~$u_i=\epsilon$. Therefore, the \smart\ strategy strictly dominates the protocol whenever $u_i>\epsilon$.
Moreover, recall that (1) the revenue per hash given by the coin in
the stable state is $\frac{w}{M\tau}$, (2) miner's $p_i$ revenue per
unit time is $\frac{w}{M\tau} m_i$, and (3) her costs are
$\mbox{\VC}_i m_i+\mbox{\FC}_i$.
Since honest miners don't mine for a loss, under our demand and supply
assumption~\cite{Wealth1776, ricardo1817}, the market powers
establish $\frac{w}{M\tau} m_i=(\mbox{\VC}_i m_i+\mbox{\FC}_i)+\epsilon$.
Assuming for simplicity~$\epsilon\rightarrow 0$, we calculate below when $u_i>0$.
\begin{align*}
	&u_i >0\\
	&\Longleftrightarrow\\
	&  \frac{w}{M\tau} m_i > \frac{M-m_i}{M} \left(\mbox{\VC}_i m_i+\mbox{\FC}_i\right) + \frac{M}{M-m_i}\mbox{FC}_i\\
	&\Longleftrightarrow\\
	& \mbox{\VC}_i m_i+\mbox{\FC}_i  >  \frac{M-m_i}{M} \left(\mbox{\VC}_i m_i+\mbox{\FC}_i\right) +\frac{M}{M-m_i}\mbox{FC}_i\\
	&\Longleftrightarrow\\
	& \frac{m_i}{M}>  \frac{\mbox{FC}_i}{\mbox{\VC}_i m_i+\mbox{\FC}_i} \cdot \frac{M}{M-m_i}\\
	&\Longleftrightarrow\\
	& \frac{m_i}{M} \cdot \frac{M-m_i}{M} >  \frac{\mbox{FC}_i}{\mbox{\VC}_i m_i+\mbox{\FC}_i}
\end{align*}
Denoting the percentage of the fixed cost out of the total costs as $y \triangleq \frac{\mbox{FC}_i}{\mbox{\VC}_i \cdot m_i+\mbox{\FC}_i}$,
and the percentage of $i$'s mining power as $x \triangleq
\frac{m_i}{M}$, we get that our smart mining attack strictly
dominates the protocol whenever $x\cdot(1-x) > y$ for $(x,y)\in
(0,1)\times (0,1)$. Figure~\ref{fig:(1-x)x} illustrates in which costs
structure the \smart\ attack dominates honest mining.
As an example, when the fixed costs are 10\% of the total costs, having 12\% of the mining power suffices to create excess profit using \smart.

\subsection{Smarter Mining}
\label{sec:smarter}
\begin{figure}[h]
    \centering
        \includegraphics[width=6.3in]{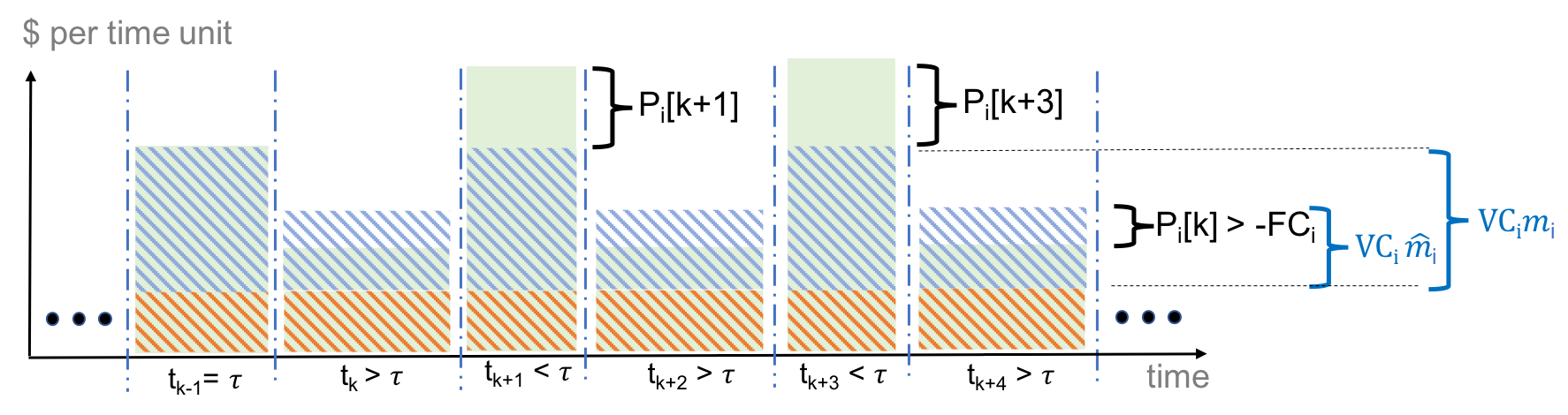}
        \caption{Smarter mining strategy for a party $p_i$.
        By mining with part of its mining power ($0 < \hat{m_i} < m_i$)
        during low revenue epochs, $p_i$ is able to maximize its
        profit by optimizing the balance between losing in low revenue
        epochs and gaining in high revenue epochs.}
    \label{fig:smarter}
\end{figure}

Having the basic \smart\ explained we can now improve it by fine tuning.
When we expand the miner's strategy space to include strategies where $0\le \hat{m_i}[k]\le m_i$ (that is, not an all or nothing but a combined strategy), the miner can achieve even greater profits by deviating from honest mining.
Moreover, attacking profits the miner in many more scenarios, both with higher fixed costs percentages and with less mining power required.
For these reasons we denote the fine-tuned attack as   \emph{smarter mining}.
Smarter mining stems from the observation that all mining operations profit in the high revenue epochs regardless of their participation in the attack, and only the idle mining power bares a loss in the low revenue epochs.
In particular, an honest miner profits from the attack without incurring any costs.
Thus, a smarter attacking miner might choose to optimize her profits by slightly reducing her excess profits in an HRE in exchange for a higher reduction of her LRE losses.
A miner that optimizes her~$\hat{m_i}[k]$ in a smarter way than~$\hat{m_i}[k]\in\{0, m_i \}$ may therefore enjoy higher average profits.
An illustration of the intuition behind smarter mining appears
in Figure~\ref{fig:smarter}.

\paragraph{Expected Profit Analysis.} The analysis is similar to
Section~\ref{sec:smart} only that in epochs~$\{ep_k, ep_{k+2},...\}$,
miner~$p_i$ mines with~$\hat{m_i}[k]$ mining power and in epochs~$\{ep_{k+1}, ep_{k+3},...\}$, it mines with its full power $m_i$.
\label{sec:Analysis}
Denote $\dmi \triangleq m_i - \hat{m_i}[k]$, that is, the amount of $p_i$'s idle mining power during epochs~$\{ep_k, ep_{k+2},...\}$, and we have the following:

\paragraph{Epoch~$k$.}
\begin{align*}
t_{k-1}&=\tau \\
    H_k&= M\tau \\
    t_k&=\frac{H_k}{M-\dmi}
    =\frac{\tau}{1-\frac{\dmi}{M}}\\
\mbox{\RpH}_{k} &= \frac{w}{H_{k}}
				  =\frac{w}{M\cdot\tau}
				  =\frac{1}{m_i}\left(\mbox{\VC}_i m_i + \mbox{\FC}_i\right)\\
 P_i[k]&= \mbox{\RpH}_{k} \hat{m_i}[k] - \left(\mbox{\VC}_i  \hat{m_i}[k]+\mbox{\FC}_i\right)
 	   = \left(\frac{\hat{m_i}[k]}{m_i} -1\right) \mbox{\FC}_i
\end{align*}

\paragraph{Epoch~$k+1$.} Since $t_k=\frac{\tau}{1-\frac{\dmi}{M}}$, the difficulty adjustment mechanism reduces the difficulty which in turn increases the reward per hash and correspondingly also the profit per time unit. The resulting values are:
\begin{align*}
H_{k+1}&= \frac{\tau}{t_k} H_{k}
	   = (M-\dmi) \tau \\
t_{k+1}&= \frac{H_{k+1}}{M}
	   = \frac{M-\dmi}{M} \tau \\
\mbox{\RpH}_{k+1} &= \frac{w}{H_{k+1}}
				  = \frac{w}{(M-\dmi)\tau}
				  = \frac{M}{M-\dmi}\mbox{\RpH}_{k}
				  = \frac{1}{m_i}\cdot\frac{M}{M-\dmi}\left(\mbox{\VC}_i m_i + \mbox{\FC}_i\right)\\	
P_i[k+1]&= m_i\mbox{\RpH}_{k+1}  - \left(\mbox{\VC}_i m_i+\mbox{\FC}_i\right)
	    = \frac{\dmi}{M-\dmi}\left(\mbox{\VC}_i  m_i+\mbox{\FC}_i\right)
\end{align*}

\paragraph{Epochs~$\ge k+2$.} Since $t_{k+1}=\frac{M-\dmi}{M} \tau$ and $H_{k+1}= (M-\dmi) \tau$, we have in $ep_{k+2}$ that $H_{k+2} = M\tau$ as in $ep_k$. The rest is identical to $ep_k$.
Inductively, from here on epochs $\{ep_{k+2}, ep_{k+4},...\}$ result in the same values as in $ep_k$, and epochs $\{ep_{k+3}, ep_{k+5},...\}$ result in the same values as in $ep_{k+1}$.

\paragraph{}As a result, if $p_i$ employs the \smarter\ strategy, her profit averaged over time as defined by the utility function would be:
\begin{align*}
u_i &\triangleq \lim_{K \to \infty} \frac{\Sigma_{k'=1}^K P_i[k']\cdot t_{k'}}{\Sigma_{k'=1}^K t_{k'}}\\
&= \lim_{K \to \infty} \left[ \frac{\Sigma_{k'=1}^K (P_i[2k'-1]\cdot t_{2k'-1}) +
	\Sigma_{k'=1}^K (P_i[2k']\cdot t_{2k'})}{\Sigma_{k'=1}^K t_{k'}} \right] \\[3ex]
&= \frac{P_i[k+1]\cdot t_{k+1} + P_i[k]\cdot t_{k}} {t_{k+1} + t_k}\\
&= \frac{\overbrace{\frac{\dmi}{M-\dmi} \left(\mbox{\VC}_i  m_i+\mbox{\FC}_i\right)}^{P_i[k+1]}  \cdot \overbrace{\frac{M-\dmi}{M} \tau}^{t_{k+1}} +
	\overbrace{\left(\frac{\hat{m_i}[k]}{m_i} -1\right) \mbox{\FC}_i}^{P_i[k]} \cdot \overbrace{\frac{\tau}{1-\frac{\dmi}{M}}}^{t_{k}} }
	{\frac{M-\dmi}{M} \cdot \tau + \frac{\tau}{1-\frac{\dmi}{M}}}
\\[3ex]
&= \frac{\frac{\dmi}{M}\left(\mbox{\VC}_i  m_i+\mbox{\FC}_i\right) -
	\frac{\dmi}{m_i} \cdot \frac{M}{M-\dmi}  \mbox{\FC}_i }
{\frac{M-\dmi}{M}  + \frac{M}{M-\dmi}}
\end{align*}

\begin{figure}[h!]
\centering
	\includegraphics[width=0.6\linewidth]{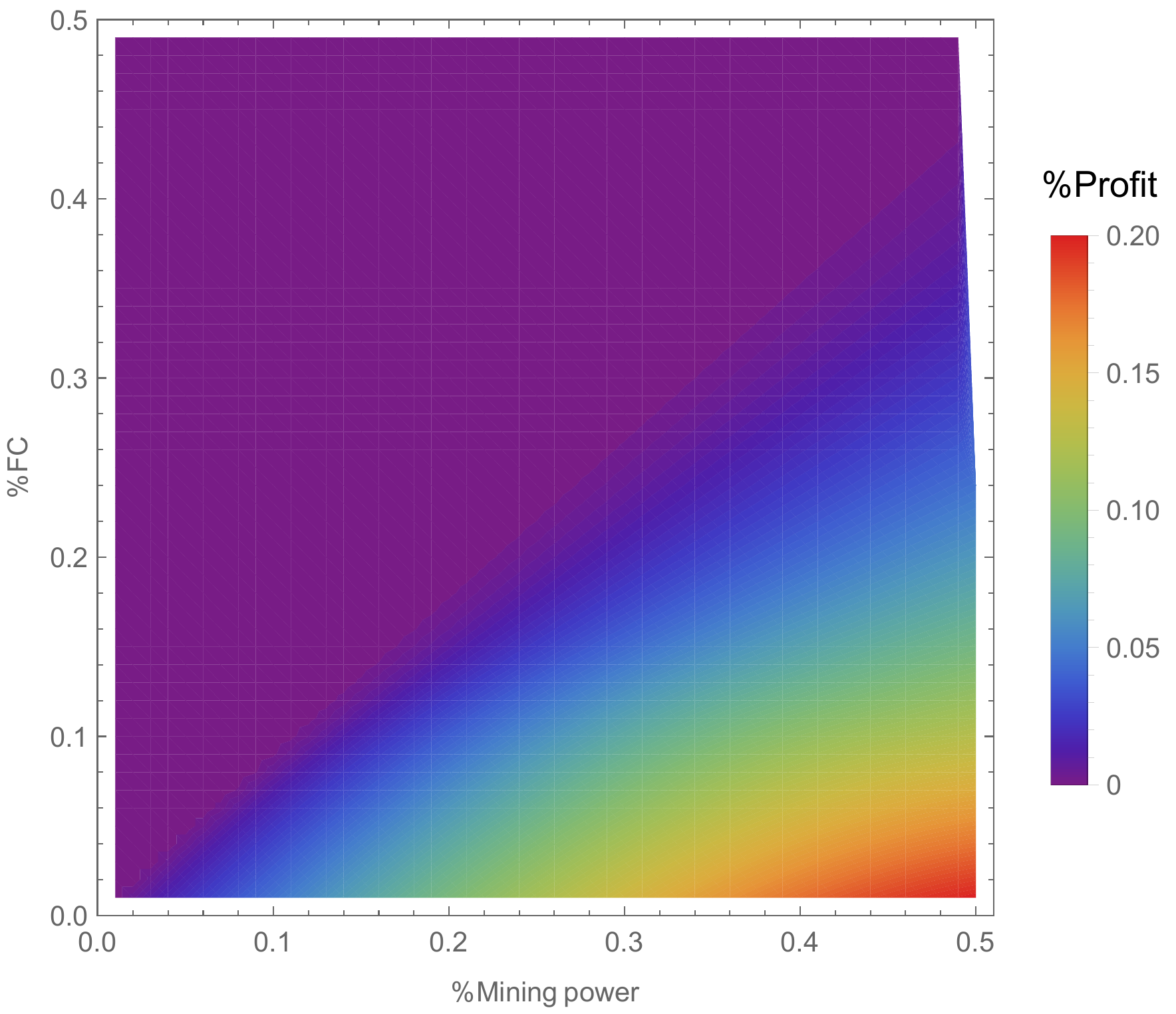}
	\caption{\smarter\ Return On Investment.
	For every~$\epsilon$, if \mbox{\textbf{\%Profit} $>\frac{\epsilon}{\VC _i  m_i+\FC _i}$}, than \smarter\ strictly dominates honest mining.
}
	\label{fig:heatMap50}
\end{figure}

Figure~\ref{fig:heatMap50} shows the strong potential of the proposed
strategy.
The colors represent profits as the percentage of the total cost ($\frac{u_i}{\mbox{\VC}_i  m_i+\mbox{\FC}_i}-1$), $x$ axis the mining power ($\frac{m_i}{M}$), and $y$ axis the costs structure ($\frac{\mbox{\FC}_i}{\mbox{\VC}_i  m_i+\mbox{\FC}_i}$).
Unlike selfish mining~\cite{SelfishMining,OptimalSelfishMining}, we can see that there are many scenarios in which \smarter\ is profitable already for a very small miner.
Furthermore, the expected profits yield a reasonable return on investment that makes \smarter\ a viable economic strategy.

\section{Coin Security and Other Parties Best Response}
\label{sec:BestResp}
As we have shown in Section~\ref{sec:RM}, the desired equilibrium of
the coin, in which the participants invest all of their power to
secure the coin (mining with full capacity), is actually not an
equilibrium in many cases.
In an abundant of realistic scenarios, a miner~$p_i$
has a strongly dominating strategy over honest mining.
One might hope, however, that the response of the other participants
in the system may somehow balance the negative effect of~$p_i$
deviating from the protocol. Unfortunately, this does not happen with
smart/\smarter.
Unlike other strategies (e.g. selfish mining \cite{SelfishMining,OptimalSelfishMining}), in our strategy~$p_i$'s deviation from the protocol does not harm the rest of the miners, on the contrary, it benefits them.

Prior to the attack, a miner~$p_j\ne p_i$ that acts honestly receives the income~$m_j\RpH_{k}$ per unit time.
After~$p_i$ starts the attack, $p_j$ receives the income per unit
time~$m_j\RpH_{k}$ (which she is satisfied with) in the
low revenue epochs, and she receives~$m_j\RpH_{k+1}>m_j\RpH_{k}$
in the high revenue epochs.
Thereby, $p_j$'s utility ($u_j$) is increased due to $p_i$'s attack.
Thus, even if~$p_j$ is rational and considers deviating from honest mining, she has no incentive to obstruct~$p_i$'s attack.
In fact, it can be shown in similar way to or analysis that not only
that~$p_j$ will not resist~$p_i$'s attack in any way, the only
possible strategy deviating from honest mining that might be more
profitable in specific cases for~$p_j$ is to join the attack and
increase both of their profits.



\paragraph{Coin security.}

Obviously, since less mining power is invested during low revenue
epochs in smart/\smarter, these strategies expose the coin
to under 50\% attacks.
For example, the \smart\ strategy is profitable for a miner that
controls 20\% of the total mining power with fixed costs that
constitutes up to 15\% of its total costs.
Therefore, if such miner chooses to adopt the \smart\ strategy, a bad
miner will be able to attack the coin with only 41\% of the total mining
power.
This vulnerability grows even more when we consider the other
miners.
As mentioned above, in some cases it is profitable for other
miners to join the smart/\smarter\ strategies, and leave the
low revenue epochs with even less honest mining power.

Surprisingly, the joining of new miners to the coin only makes the
security problem worse in smart/\smarter\ strategies.
For comparison, in long term analysis of selfish
mining~\cite{SelfishMining}, we get that new miners benefit from
joining the coin due to the drop in difficulty, which leads to higher
RpH.
These new miners mitigate the loss in honest mining power due
to forks created by selfish mining. 
However, this is not the case in smart/\smarter.
Obviously joining miners will join the epoch that are profitable for
them, which are the high revenue epochs (smart/\smarter\ does
not change the RpH in low level epochs, so if these epochs were not
profitable for new miners before, they remain unprofitable after
the smart/\smarter\ attack is performed).
This, in turn, will lead to a drop in the RpH in the low revenue
epochs due the difficulty adjustment, and force more miners to abandon
these epochs, leaving them with even less honest mining power.

%


\section{Discussion}
\label{sec:Disc}

The recent drop in the Bitcoin~\cite{Bitcoin} price makes mining much
less profitable than in the past, and force miners to revisit their
mining strategies.
In this paper we propose two strategies, smart and \smarter,
that, in many cases, strictly dominate honest mining.
In contrast to other known attacks that induce zero-sum games among
miners~\cite{SelfishMining}, the strategies proposed in this paper increase
miners' profits by reducing their variable costs (i.e., electricity).
However, while saving electricity power is very encouraging for the
environment, it is less so for the coin security.
The smart/\smarter\ strategies expose the coin to under
50\% attacks during low revenue epochs, and this vulnerability only
grows when new miners join the coin as a response to the increase in profit
margins induced by these strategies.

Interestingly, the smart/\smarter\ can be profitable even for miners that poses a relatively small amount of mining power, and in case the fixed cost is negligible even less than 1\% is enough.
Another point that is worth to mention is the potential of smart/\smarter\ in a system with many PoW-based coins.
Although this paper deals with a single coin system, the extension
into multiple coins only strengthen the attack strategies, since it
allows attackers to mine for other coins during low revenue epochs
and thus significantly reduce the loss in these epochs or even make
profit by joining other coins in their high revenue epochs.%
\footnote{This trivial property makes the attack somewhat unbeaten in
the practical world and thus poses a significant threat to PoW-based
cryptocurrencies.}

 
In this work we have exploited the difficulty adjustment mechanisms of PoW-based cryptocurrencies by manipulating it to our benefit.
In general, this is but a single case of trying to emulate a continues process (supply and demand adjustment) by a discrete process (difficulty update in epochs).
Similar issues are likely to arise in other mechanisms and systems, as the transformation from continues to discrete is not trivial, but does not always receive the appropriate attention during the design of a system.
We believe that areas such as blockchains, where real world systems
progress faster than rigorous analysis, provide many opportunities
for theoretical research to make meaningful contributions.

%


\bibliographystyle{plain}
\bibliography{z}

\begin{thebibliography}{10}

\bibitem{Powercompare}
Bitcoin power compare.
\newblock \url{https://powercompare.co.uk/bitcoin/}, Accessed: 2019-01-28.

\bibitem{MarketState}
Cryptocurrency market state visualization.
\newblock \url{https://coin360.io/}, Accessed: 2019-01-28.

\bibitem{spiral}
Bitcoin power compare.
\newblock
  \url{https://www.theblockcrypto.com/2018/12/04/the-bitcoin-mining-death-spiral-debate-explained/
  }, Accessed: 2019-01-30.

\bibitem{ethereum}
Ethereum foundation.
\newblock \url{https://www.ethereum.org/}, Accessed: 2019-02-07.

\bibitem{back2002hashcash}
Adam Back et~al.
\newblock Hashcash-a denial of service counter-measure, 2002.

\bibitem{bonneau2016buy}
Joseph Bonneau.
\newblock Why buy when you can rent?
\newblock In {\em International Conference on Financial Cryptography and Data
  Security}, pages 19--26. Springer, 2016.

\bibitem{carlsten2016instability}
Miles Carlsten, Harry Kalodner, S~Matthew Weinberg, and Arvind Narayanan.
\newblock On the instability of bitcoin without the block reward.
\newblock In {\em Proceedings of the ACM SIGSAC Conference on Computer and
  Communications Security}, pages 154--167. ACM, 2016.

\bibitem{decker2013information}
Christian Decker and Roger Wattenhofer.
\newblock Information propagation in the bitcoin network.
\newblock In {\em Peer-to-Peer Computing (P2P), 2013 IEEE Thirteenth
  International Conference on}, pages 1--10. IEEE, 2013.

\bibitem{MinersDilemma}
Ittay Eyal.
\newblock The miner's dilemma.
\newblock In {\em Security and Privacy (SP), 2015 IEEE Symposium on}, pages
  89--103. IEEE, 2015.

\bibitem{SelfishMining}
Ittay Eyal and Emin~Gun Sirer.
\newblock Majority is not enough: Bitcoin mining is vulnerable.
\newblock In {\em International Conference on Financial Cryptography and Data
  Security}, pages 436--454. Springer, 2014.

\bibitem{hellemans2019mining}
Tim Hellemans, Benny Van~Houdt, Daniel~S Menasche, Mandar Datar, Swapnil
  Dhamal, and Corinne Touati.
\newblock Mining competition in a multi-cryptocurrency ecosystem at the network
  edge: a congestion game approach.
\newblock {\em ACM SIGMETRICS Performance Evaluation Review}, 46(3):114--117,
  2019.

\bibitem{Johnson14}
Benjamin Johnson, Aron Laszka, Jens Grossklags, Marie Vasek, and Tyler Moore.
\newblock Game-theoretic analysis of ddos attacks against bitcoin mining pools.
\newblock In {\em International Conference on Financial Cryptography and Data
  Security}, pages 72--86. Springer, 2014.

\bibitem{MiningGames}
Aggelos Kiayias, Elias Koutsoupias, Maria Kyropoulou, and Yiannis Tselekounis.
\newblock Blockchain mining games.
\newblock In {\em Proceedings of the 2016 ACM Conference on Economics and
  Computation}, pages 365--382. ACM, 2016.

\bibitem{liao2017incentivizing}
Kevin Liao and Jonathan Katz.
\newblock Incentivizing blockchain forks via whale transactions.
\newblock In {\em International Conference on Financial Cryptography and Data
  Security}, pages 264--279. Springer, 2017.

\bibitem{Bitcoin}
Satoshi Nakamoto.
\newblock Bitcoin: A peer-to-peer electronic cash system, 2008.

\bibitem{StubbornMining}
Kartik Nayak, Srijan Kumar, Andrew Miller, and Elaine Shi.
\newblock Stubborn mining: Generalizing selfish mining and combining with an
  eclipse attack.
\newblock In {\em 2016 IEEE European Symposium on Security and Privacy
  (EuroS\&P)}, pages 305--320. IEEE, 2016.

\bibitem{ricardo1817}
David Ricardo.
\newblock On the principles of political economy and taxation.
\newblock 1817.

\bibitem{OptimalSelfishMining}
Ayelet Sapirshtein, Yonatan Sompolinsky, and Aviv Zohar.
\newblock Optimal selfish mining strategies in bitcoin.
\newblock In {\em International Conference on Financial Cryptography and Data
  Security}, pages 515--532. Springer, 2016.

\bibitem{Schrijvers16}
Okke Schrijvers, Joseph Bonneau, Dan Boneh, and Tim Roughgarden.
\newblock Incentive compatibility of bitcoin mining pool reward functions.
\newblock In {\em International Conference on Financial Cryptography and Data
  Security}, pages 477--498. Springer, 2016.

\bibitem{Wealth1776}
Adam Smith.
\newblock An inquiry into the nature and causes of the wealth of nations, 1776.

\bibitem{spiegelman2018game}
Alexander Spiegelman, Idit Keidar, and Moshe Tennenholtz.
\newblock Game of coins.
\newblock {\em arXiv preprint arXiv:1805.08979}, 2018.

\bibitem{tsabary2018gap}
Itay Tsabary and Ittay Eyal.
\newblock The gap game.
\newblock In {\em Proceedings of the 2018 ACM SIGSAC Conference on Computer and
  Communications Security}, CCS '18, pages 713--728, New York, NY, USA, 2018.
  ACM.

\end{thebibliography}

\appendix

\end{document}